\documentclass[11pt]{article}
\usepackage[margin=1in]{geometry}
\usepackage{lmodern}
\usepackage{sidecap}
\usepackage{xspace}
\usepackage{doi}
\usepackage{hyperref} 
\usepackage{tikz}
\usepackage{pgfgantt}
\usepackage{graphicx,pbox}
\usepackage{xcolor,hyperref,colortbl}
\usepackage{wrapfig}
\usepackage{enumitem}
\usepackage{gensymb}
\usepackage{comment}

\newcommand\pubnumber{ }
\newcommand\pubdate{\today}
\newcommand\pubblock{\rightline{\begin{tabular}{l} \pubnumber\\
          \pubdate \end{tabular}}}

\def\Title#1{\begin{center} {\Large #1 } \end{center}}
\def\Author#1{\begin{center}{ \sc #1} \end{center}}

\newcommand\snowmass{\begin{center}\rule[-0.2in]{\hsize}{0.01in}\\\rule{\hsize}{0.01in}\\
\vskip 0.1in Submitted to the  Proceedings of the US Community Study\\ 
on the Future of Particle Physics (Snowmass 2021)\\ 
\rule{\hsize}{0.01in}\\\rule[+0.2in]{\hsize}{0.01in} \end{center}}
\begin{document}
\begin{titlepage}
\snowmass
\pubblock

\Title{Advanced RF Sources R$\&$D for Economical Future Colliders}


\Author{Brandon Weatherford$^{1}$, Emilio A. Nanni$^{1}$,  and  Sami Tantawi$^{1}$}

\noindent
$^{1}${SLAC National Accelerator Laboratory, Stanford University}\\



\end{titlepage}
\newpage

\tableofcontents
\newpage

\addcontentsline{toc}{section}{Executive Summary}
\section*{Executive Summary}
The high power RF system will be a significant budgetary driver for any future collider. An order-of-magnitude improvement in cost/capability is needed, and as a result, a robust R$\&$D program in next-generation, economical RF sources is essential. In this paper, we discuss the challenges and opportunities that arise from advancing the state of the art in these devices.  Specifically, research initiatives in new circuit topologies, advanced manufacturing techniques, and novel alternatives to conventional RF source components are discussed.

\newpage

\section{Introduction} 

In any future large-scale, normal conducting RF accelerator facility, the capital and operational costs of the RF power chain – the high power RF amplifiers (klystrons) and the high voltage modulators for powering them – are substantial\cite{aryshev2022international,bai2021c,dasu2022strategy}.  Reasonable budgeting models for a TeV-scale collider predict that the high power RF system alone would cost more than the construction of the accelerator itself and the tunnel required to house it; on the order of \$10B.  These estimates assume the use of existing RF amplifier and modulator technologies at a combined $\sim$\$20/peak kW.  As was stated in the 2017 DOE Radiofrequency Accelerator R$\&$D Strategy Report: 

\begin{quote}
    "When pushing high gradient or intensity limits, RF sources become the leading cost driver for normal conducting accelerators and a significant cost driver for superconducting accelerators. Only with innovative concepts for designing and building RF sources will dramatic reduction in cost and increased efficiency be achieved."
\end{quote}

This same report suggests a target of $\$$2/peak kW - at least ten times less expensive than what is commercially available today.  Whether klystrons, magnetrons, or other existing RF sources are employed, the relative differences in their costs – even considering trade-offs with respect to peak vs. average power, or operational lifetime – are negligible when compared to the order of magnitude reduction in cost that is needed.

\section{Research Initiatives}

RF amplifiers are often thought of as “known quantities,” and while there is an abundance of research activity in accelerating structures, R$\&$D in high power RF sources is relatively uncommon. Maybe it is assumed that industry will solve the cost/capability challenge for RF power – but the prospects for this are grim indeed. Unfortunately, there are not many commercially viable uses for megawatt-class RF amplifiers, and the devices that do exist are usually custom-designed for a specific application. It is unreasonable to expect that an industrial supplier will invest their own time and money to reduce cost in anticipation of a single-use system that may (or may not) be assembled decades in the future. Because of the long time frame, high technical risk, and undefined initial requirements associated with a next generation collider, any reasonable business plan would “price in” the impact of these uncertainties – so when a new facility is proposed, the RF power system will be prohibitively expensive. Support for high-risk research in RF sources is desperately needed if we are to realize the order of magnitude improvement in cost-capability that is truly required. Such an effort must be led by government labs and academia because these institutions can tolerate the long term risk of this effort. 

An example of a successful lab-led R$\&$D initiative in high power RF sources has been the High Efficiency International Klystron Activity (HEIKA), established by CERN in 2014. This effort has supported worldwide collaboration to understand in detail how to optimize the electrical design of high power klystrons for maximum DC-to-RF efficiency. As a result of this work, new klystron design methods such as the Core Oscillation Method (COM) and Core Stabilization Method (CSM) were established. New designs for COM-based 8 MW and 50 MW X-band klystron prototypes have been completed by CERN and industrial partners.\cite{cai2021XBandKly} In addition, a publicly available 2D klystron simulation code called KlyC was successfully developed, benchmarked, and made available to the RF source community.\cite{cai2019klyc}  The tools and design improvements arising from the HEIKA collaboration will be helpful in developing high efficiency (and therefore lower operational cost) prototypes for any future demonstration-scale collider, but more work still must be done to reduce the upfront construction costs of HPRF sources. 

\begin{figure}[!ht]
    \centering
    \includegraphics[width=0.49\textwidth]{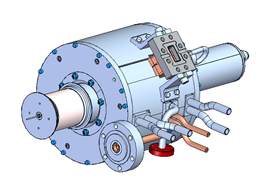}
    \hfill
    \includegraphics[width=0.49\textwidth]{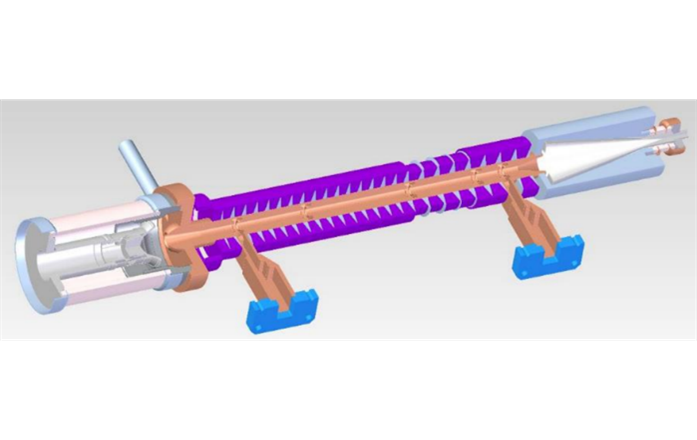}
    \caption{Examples of compact, low-voltage klystron designs for commercially compact accelerator systems, which could be standardized and produced at large volume. [B. Weatherford and S. Tantawi]}
    \label{fig:klystrinos}
\end{figure}

In the near term, research institutions should also identify and aggressively pursue new applications of RF sources and accelerator systems in the commercial, defense, and medical sectors; and to the degree possible, develop broadly useful RF source topologies that could have a need in high volume production. As one example, compact low-voltage klystron amplifiers are being developed for use in multiple linac-based radiography systems (see Figure \ref{fig:klystrinos}). If such a “building block” RF source can be optimized for use in small and large accelerator systems alike, and standardized as much as possible, then commercial opportunities and real competition between suppliers would drive significant cost reductions. Compare this to the approach of using custom-designed RF sources for a single facility, and the path to improved cost/capability is clear. 

Longer term, fundamental and exploratory research dedicated to RF sources and their components is essential for more than marginal improvements. The most reasonable approach involves optimizing the complete RF power chain, which naturally leads to using lower voltage modulators made from mass-produced commercially available components, which are simpler and require less infrastructure and maintenance.  Then, new RF sources are needed which can operate efficiently at low voltage and high current. Multiple-beam amplifiers leverage this concept, but this scaling approach can add complexity and does not really solve the fundamental problem - breaking the tradeoff between efficiency and perveance that is inherent in conventional linear-beam devices. Reconsidering RF sources in this way raises several interesting fundamental physics and engineering challenges. 

One approach involves using multi-dimensional electron beams, Figure \ref{ig:multidimRF}, allowing for higher current densities at a given voltage. This relaxes requirements on the beam focusing magnets and could potentially minimize the impact of space charge on RF efficiency. In another approach, the trajectory of a monoenergetic beam is modulated and interacts with a higher order mode in the output cavity; this could be promising for high efficiency amplification or frequency multiplication. 

\begin{figure}[!ht]
    \centering
    \includegraphics[width=0.6\textwidth]{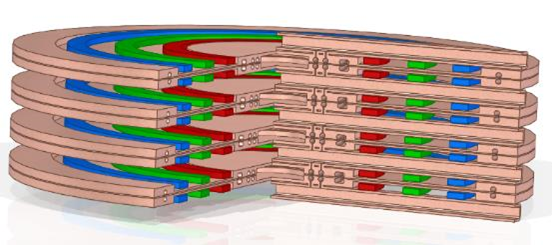}
    \includegraphics[width=0.9\textwidth]{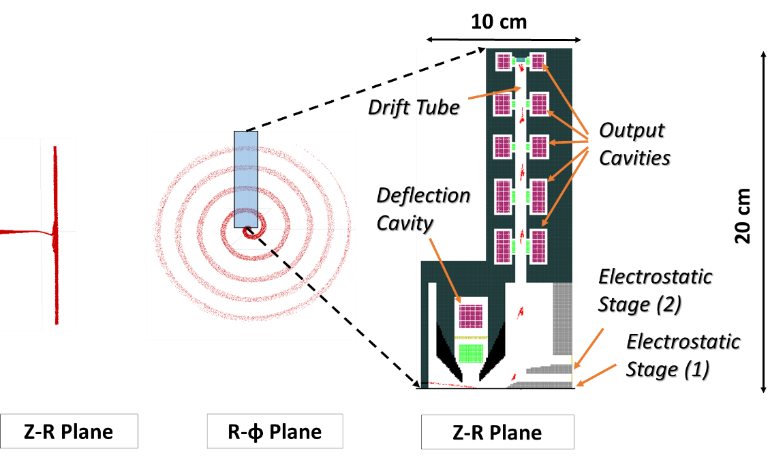}
    \caption{RF source topologies utilizing multi-dimensional beams. Radial klystron array (top) [S. Tantawi]; Deflected beam amplifier (bottom) [M. Franzi and S. Tantawi]}
    \label{fig:multidimRF}
\end{figure}

Just as the chain of modulator and RF source needs to be optimized holistically, the same is true of the components which comprise the RF source itself. The most expensive parts of existing klystron builds tend to be the electron source, focusing magnet, and the assembly labor (via multiple brazing steps). These pain points should all be attacked in parallel because a breakthrough in any one area will create new opportunities in the others. For example, if a high current density electron source can be generated that is less sensitive to impurities than thermionic cathodes, then perhaps the source can operate with a lower quality vacuum, and the assembly process can be simplified. Similarly, advances in microfabrication or additive manufacturing could enable new field emitters or focusing magnet topologies, respectively. 

\begin{figure}[!ht]
    \centering
    \includegraphics[width=0.11\textwidth]{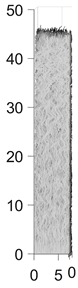}
    \hspace{0.1\textwidth}
    \includegraphics[width=0.45\textwidth]{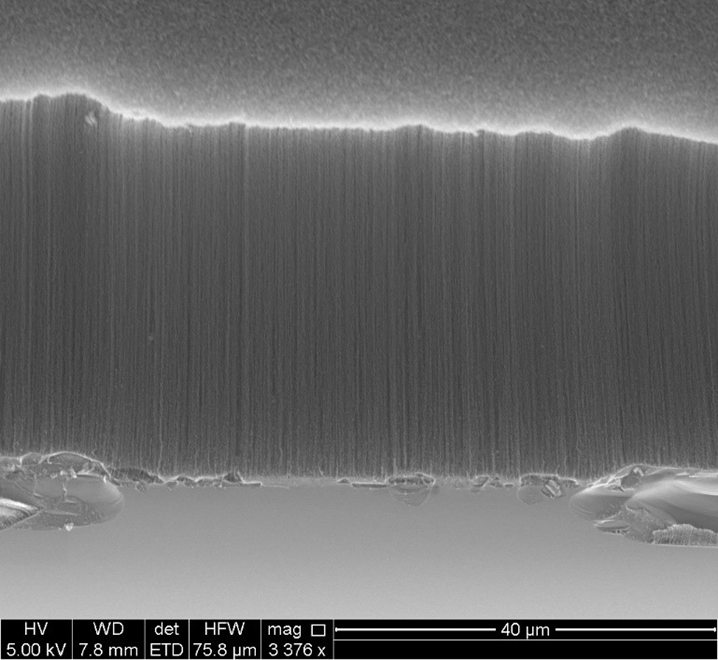}
    \caption{Growth simulation (left) and SEM images (right) of carbon nanotube forest samples for field emitter characterization studies. [M. Maschmann, S. Kovaleski, and B. Weatherford]}
    \label{fig:CNT}
\end{figure}

There are many interesting possibilities for a robust RF source research program. Novel alternatives to thermionic cathodes are needed, especially if they are suitable for massively parallel systems. Plasma cathode electron sources can deliver up to hundreds of Amperes of electron current at hundreds of Volts, and can operate at substantial pressures, partially mitigating the need for pristine vacuum environments. Advances in the control of carbon nanotube growth and patterning processes (Figure \ref{fig:CNT}), along with their high resilience to ion bombardment, present another possibility, as do microfabricated field emitter arrays. Beam focusing in RF sources is another major challenge – with magnets for a single klystron costing tens of $\$$k themselves, alternative focusing approaches for low voltage beams should be investigated. Potential areas of interest could include multiple stage electrostatic focusing of low voltage beams, self-focusing of high current beams in a plasma background, and hybrid approaches of combined electrostatic and magnetic focusing. These efforts would require significant computational and experimental work but raise many interesting basic physics questions, and in some cases, could be synergistic with R$\&$D efforts in the plasma wakefield accelerator community. 

\begin{figure}[!ht]
    \centering
    \includegraphics[width=1\textwidth]{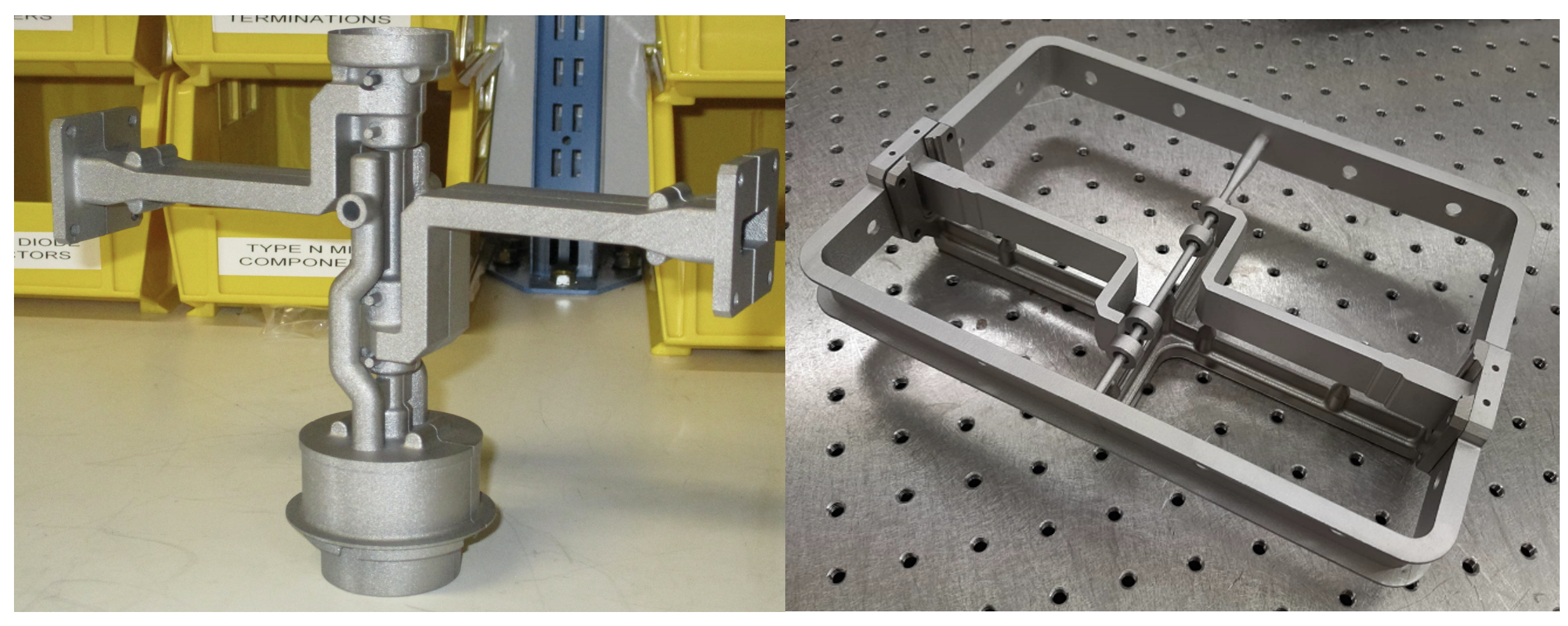}
    \caption{Klystron body produced by direct metal laser sintering (left); Mandrel (negative space) of klystron RF circuit (right). [J. Merrick, C. Wehner, and E. Nanni]}
    \label{fig:AM}
\end{figure}

Additive manufacturing is a promising development for minimizing the touch labor required to assemble an RF source, and the availability of commercial additive processes is growing rapidly. Developing inexpensive additive processes specifically for the niche of vacuum devices is a difficult undertaking. However, widely available additive manufacturing processes are constantly evolving, and many could be suitable for RF source fabrication, but they have not been fully characterized with respect to RF losses, high vacuum, high voltage, etc. In certain subsystems of an RF amplifier, particularly in klystrons where RF ohmic losses are only appreciable near the tube output, or in low-duty operation, additive manufacturing could be feasible (see Figure \ref{fig:AM}). When existing additive processes are not suitable, fundamental research in novel additive manufacturing processes should be supported as well.

\section{Conclusion}

In summary, the high cost of RF power can be a major budgetary constraint for any new high energy physics facility. Although it is often assumed that RF sources are “established,” the truth is the current state of the art is nowhere near what the high energy physics community needs regarding cost/capability. Solving this enormous physics and engineering problem will require significant support for both computational and experimental R$\&$D efforts in RF sources\cite{nanni2022c}. A real solution demands that we fundamentally re-imagine RF source topologies and rigorously attack the many exciting basic questions that arise from this challenge.

\section{Acknowledgements}

The work is supported in part by Department of Energy Contract DE-AC02-76SF00515.

\addcontentsline{toc}{section}{Bibliography}

\bibliographystyle{atlasnote}
\bibliography{bibliography.bib}

\end{document}